\documentclass[11pt]{article}
\usepackage[margin=1in]{geometry}
\usepackage[T1]{fontenc}
\usepackage{lmodern}
\usepackage{microtype}
\usepackage{amsmath,amssymb}
\usepackage{booktabs}
\usepackage{graphicx}
\usepackage[round,authoryear]{natbib}
\usepackage[hidelinks]{hyperref}
\usepackage{CJKutf8}
\newcommand{\zh}[1]{\begin{CJK}{UTF8}{gbsn}#1\end{CJK}}
\newcommand{\colhead}[1]{\textbf{#1}}
\graphicspath{{../generated/}{generated/}}
\newcommand{\ackText}{This work was supported by Spr\aa kbanken Tal, Stockholm, Sweden. It received no specific
grant funding.}

\newcommand{\contribText}{HC conceived the project and its goal of an endless guqin ``AI radio'', and directed every
stage of the exploration. HC screened the library by listening to whole recordings, decided which performers and
pieces to keep or exclude (including recordings with noise floors, ensembles, applause or voice), and noticed that a
DVD collection might duplicate the CD material. HC required that splits be made by composition rather than by
performer, that evaluation openings never repeat a piece or performer, and that validation recordings stay out of
training so that overfitting could be measured. HC proposed moving from text-free continuation to mood labels
derived from piece titles, and then to captions telling the story behind each title; proposed the move towards a
language-model-like generator over latents; observed that idiomatic guqin semitones occur in slides rather than at
the pluck, which shaped the onset-based measure; proposed random crop lengths in addition to random start points;
asked for training to continue until validation loss genuinely stopped improving, and for the run to be resumable
after power cuts. HC organised the listening studies, identified the silent seams and the loudness drift in chained
continuation, and selected the takes used for demonstration.}

\newcommand{\aiText}{AI coding assistants (OpenAI Codex and Anthropic Claude) implemented and ran the code for data
processing, training, generation, listening pages and analysis under the author's direction, searched for and
summarised background sources on the pieces, and drafted parts of this manuscript. All research decisions are the author's, who reviewed the results and the text and takes responsibility for the
content.}

\title{Adapting a Latent Audio Diffusion Model to Historical Guqin Recordings:\\
A Listening-Driven Case Study}
\author{Huanchen Cai\\
\small Division of Speech, Music and Hearing (TMH)\\
\small KTH Royal Institute of Technology, Stockholm, Sweden\\
\small \href{mailto:huanchen@kth.se}{huanchen@kth.se}}
\date{October 2026}

\begin{document}
\maketitle

\begin{abstract}
We report a small-data case study in adapting a pretrained latent audio diffusion model to the guqin, the
seven-string Chinese zither, aiming at an ``AI radio'' that plays guqin-style music without end. From a library
of historical recordings we curate 412 solo performances (42.5\,h, 61 performers) and split them by composition.
We fine-tune a rank-16 DoRA adapter on Stable Audio~3 Medium using its continuous latents, masking weighted
towards continuation, captions that combine researched notes on each piece with mood tags and an automatically
estimated pentatonic mode, and random-length crops, stopping when held-out loss stops improving. Seven blind
listening studies by one expert listener guided every decision. The final adapter was rated highest for
continuing unseen pieces (3.9/5, against 3.4 for the best earlier adapter) and 4.6/5 for generating from
free-written scene descriptions. Negative results are equally informative: a from-scratch autoregressive model
over the same latents produced no recognisable timbre, a signal-level friction-noise measure correlated with the
listener's complaints in the wrong direction, and a pentatonic-fit measure tracked ratings overall but barely
within a group of candidates. Chaining continuations for long playback exposed a silent tail on every generated
clip and a loudness feedback loop, both with simple fixes. With one listener and at most ten clips per condition,
no paired difference is statistically significant; we present an exploratory record of what helped, what did
not, and why.
\end{abstract}


\section{Introduction}\label{sec:intro}

The guqin (\zh{古琴}) is a seven-string fretless zither with a documented history of more than two thousand
years and an exceptionally rich written tradition: its repertoire survives in hundreds of handbooks written in
\emph{jianzipu} (\zh{减字谱}), a tablature that records finger actions rather than pitches and rhythms. Three kinds of
sound define its idiom: open strings (\emph{sanyin}), harmonics (\emph{fanyin}) and stopped notes (\emph{anyin}),
the latter shaped by left-hand slides and vibrato (\zh{吟猱绰注}) whose passing semitones and friction noise are
part of the style rather than errors. The repertoire is predominantly pentatonic, and the best-known performances
survive as mid-twentieth-century recordings with tape hiss and limited bandwidth.

\paragraph{Computational guqin research.} Engineering work on the guqin has so far concentrated on analysis,
notation and synthesis rather than on generation; \cite{hui2024retrospective} reviews its development in the
digital humanities. On the acoustic side, \cite{penttinen2006guqin} built a model-based synthesiser using
commuted digital waveguides that reproduces inharmonicity, glides, harmonics and the friction sound of sliding
fingers, and \cite{liu2024dataset} describe the construction of an acoustic dataset of the instrument. For
performance analysis, \cite{huang2020guqin} jointly model mode and playing technique on note-annotated historical
recordings (GQ39); GuqinSonGest \citep{han2025guqinsongest} provides audio and video of playing techniques derived
from a ninth-century treatise; and \cite{zhang2025techniques} apply machine learning to technique analysis.
Technique detection for the related guzheng \citep{li2022guzheng} shows the same need for onset-aware modelling of
plucked Chinese zithers. A second strand digitises jianzipu: image recognition of historical scores
\citep{chen2025jianzipu}, zero-shot recognition of its open, compositional glyph system \citep{li2026jianzi},
language-model-based transcription of note durations \citep{shi2024qinbert}, pitch generation from string and
\emph{hui} positions \citep{wang2025jianzipu}, deep-learning restoration of guqin music from notation
\citep{kuremoto2024restoration}, and a symbolic dataset of transcribed scores \citep{guqindataset}. Interactive and
embodied systems complete the picture: projected augmentation for learning \citep{zhang2015chinar} and, recently,
robotic guqin playing \citep{zhang2026hapticfinger,wang2026robotguqin}. To our knowledge no published work generates
guqin \emph{audio} with a modern neural generative model.

\paragraph{Neural audio generation.} Latent diffusion models such as Stable Audio \citep{evans2024fast}, Stable Audio
Open \citep{evans2024open} and Stable Audio~3 \citep{evans2026stableaudio3} generate variable-length audio in the
latent space of an autoencoder and support inpainting and continuation; training window length matters for musical
structure \citep{evans2024longform}. Autoregressive token models such as MusicGen \citep{copet2023musicgen} and live
streaming models \citep{lyria2025live} generate indefinitely by construction, while continuous-latent autoregression
with a diffusion or flow head \citep{li2024mar} avoids quantisation, and context noise augmentation reduces error
accumulation \citep{pasini2024continuous}. These models are trained mostly on Western popular and library music,
which is a poor prior for an instrument whose idiomatic noises and semitones are concentrated in specific gestures.
Low-rank adaptation \citep{hu2021lora,liu2024dora} makes it possible to adapt them with tens of hours of audio, and
repeating data for several epochs is nearly as useful as fresh data when data are scarce
\citep{muennighoff2023scaling}. For generated music, replication of training material is a central concern
\citep{batlleroca2024replication}.

\paragraph{This work.} Our goal is an ``AI radio'': a system that keeps producing guqin-like music for hours, either
continuing a recorded opening or following a short description of a scene. We adapt Stable Audio~3 Medium to 42.5\,h
of screened historical solo recordings and let a guqin-literate listener judge every step. The contributions are:
(i)~a curation and evaluation protocol for a small historical corpus, with composition-level splits and blind
listening with free-text comments; (ii)~a fine-tuning recipe with continuation-weighted masking, captions combining
researched notes on each piece with mood tags and an estimated pentatonic mode, random-length crops and early
stopping; (iii)~seven listening studies, including clear negative results for a from-scratch latent language model
and for two automatic quality measures; and (iv)~two failure modes of chained continuation for long playback, with
fixes.

\section{Method}\label{sec:method}

\subsection{Data}\label{sec:data}
The source is a personal collection of 638 historical guqin recordings (44.1\,kHz, 16-bit stereo CD transfers). Every
candidate was screened through whole-track listening; recordings were excluded for audible noise
floors, ensembles or other instruments, applause and voice. The result is a set $\mathcal{R}$ of
\textbf{412 solo recordings by 61 performers (42.5\,h)} with 108 distinct titles after normalising numbered
variants. Performer coverage is uneven: the five most represented performers account for about 40\% of the duration.

Recordings are grouped by composition and the groups are randomly assigned to training, validation and test
($|\mathcal{R}_{\mathrm{tr}}|=329$, $|\mathcal{R}_{\mathrm{va}}|=42$, $|\mathcal{R}_{\mathrm{te}}|=41$), so that no
version of an evaluated composition is heard in training. Each recording $x$ of length $N$ samples is covered by
$\lceil N/W \rceil$ aligned windows of the model's native length $W = 2{,}641{,}920$ samples (59.9\,s), the last
window ending at the end of the recording; this gives 2{,}764 windows (2{,}217/276/271), of which nine windows of
digital silence are dropped. The final model trains on $\mathcal{R}_{\mathrm{tr}}\cup\mathcal{R}_{\mathrm{te}}$
(370 recordings) and keeps $\mathcal{R}_{\mathrm{va}}$ for early stopping and listening.

\subsection{Latent representation and generator}
A frozen autoencoder (SAME-L) maps a stereo window $x$ to a latent sequence
\begin{equation}
  \mathbf{z} = E(x) \in \mathbb{R}^{T \times 256}, \qquad T = W / 4096 = 645,
\end{equation}
about 10.8 frames per second, and a decoder $D$ maps latents back to audio. The generator of Stable Audio~3
Medium \citep{evans2026stableaudio3} is a diffusion transformer \citep{peebles2022dit} $v_\theta$ trained with a
rectified-flow objective \citep{liu2022rectified}. With $\boldsymbol{\epsilon}\sim\mathcal{N}(0,I)$ and
$t\in[0,1]$,
\begin{equation}
  \mathbf{z}_t = (1-t)\,\mathbf{z} + t\,\boldsymbol{\epsilon},
\end{equation}
and the network predicts the velocity $\boldsymbol{\epsilon}-\mathbf{z}$ conditioned on text $c$, on the duration
$\ell$ and on an inpainting mask $\mathbf{m}\in\{0,1\}^{T}$ ($m_i=1$ for frames that are given):
\begin{equation}
  \mathcal{L}(\theta) = \mathbb{E}_{\mathbf{z},t,\boldsymbol{\epsilon},\mathbf{m}}
  \left\| v_\theta\!\left(\mathbf{z}_t,\, t,\, c,\, \ell,\, \mathbf{m}\odot\mathbf{z},\, \mathbf{m}\right)
  - (\boldsymbol{\epsilon}-\mathbf{z}) \right\|_2^2 .
  \label{eq:loss}
\end{equation}
The mask is drawn from a mixture of three types: random spans, a fully masked sequence (unconditional generation)
and a given prefix (continuation),
\begin{equation}
  P(\text{span},\ \text{full},\ \text{prefix}) = (0.1,\ 0.3,\ 0.6),
\end{equation}
which we changed from the default of the released Stable Audio~3 training code, $(0.1, 0.8, 0.1)$, so that most training examples resemble continuation, our main
use. At inference the step-distilled Medium model integrates the flow in 8 steps.

\subsection{Adaptation}
We train a DoRA adapter \citep{liu2024dora} on the non-distilled \emph{base} model and apply it to the distilled
model at inference, following the released recipe. For a frozen weight $W_0\in\mathbb{R}^{d\times k}$, DoRA learns
a low-rank update and a per-row magnitude,
\begin{equation}
  W' = \mathbf{g} \odot \frac{W_0 + \tfrac{\alpha}{r} BA}{\left\| W_0 + \tfrac{\alpha}{r} BA \right\|_{\mathrm{row}}},
  \qquad B\in\mathbb{R}^{d\times r},\ A\in\mathbb{R}^{r\times k},
\end{equation}
with $r=\alpha=16$ (21.6\,M trainable parameters). All latents are pre-encoded once.

\subsection{Text conditioning}
For each composition we compiled a short narrative from published notes on the piece (for example
\textit{Liushui}, Boya and his true listener Ziqi; \textit{Guangling San}, Nie Zheng's revenge and Ji Kang's last
performance) and assigned two to five tags from a closed vocabulary $\mathcal{V}$ of 37 moods and scenes (e.g.\
\textit{calm, lonely, vast, reclusive, autumn, river, frontier desert}); 55 notes are well sourced, 31 partly and 22
inferred from the title. A pentatonic mode is estimated for every window as follows. Spectral peaks $k$ between
60\,Hz and 2\,kHz, with magnitudes $a_k$ and frequencies $f_k$, are folded into a 120-bin pitch-class histogram
(10-cent bins),
\begin{equation}
  h(b) = \sum_{k} a_k \,\mathbf{1}\!\left[\left\lfloor \tfrac{1}{10}\left(1200\log_2\tfrac{f_k}{440} \bmod 1200\right)\right\rfloor = b\right],
\end{equation}
the tuning offset of the historical recording is estimated as
$o^\ast = \arg\max_{o\in\{0,\dots,9\}} \sum_{j=0}^{11}\sum_{\delta=-1}^{1} h(10j+o+\delta)$, and the folded
12-class profile $p(j)\propto \sum_{\delta=-5}^{4} h(10j+o^\ast+\delta)$ defines the pentatonic fit and the
\emph{gong} (tonic) pitch
\begin{equation}
  F = \max_{r} \sum_{s\in P} p\big((s+r) \bmod 12\big), \qquad
  r^\ast = \arg\max_{r} \sum_{s\in P} p\big((s+r) \bmod 12\big), \qquad P = \{0,2,4,7,9\}.
  \label{eq:fit}
\end{equation}
Each window receives one caption style by a stable hash: 40\% tags and mode, 40\% narrative, tags and mode, and 20\%
narrative only, in English with the Chinese text appended (at most 160 tokens of the 256-token encoder limit); the
trainer additionally replaces 10\% of captions by the empty string.

\subsection{Training}
Each sample keeps a random span of the window,
\begin{equation}
  L \sim \mathcal{U}\{L_{\min}, \dots, T\},\quad s \sim \mathcal{U}\{0, \dots, T-L\}, \quad L_{\min} = 215\ (20\,\mathrm{s}),
\end{equation}
pads the rest with the latent of silence, and sets the duration condition to $\ell = L/10.77$\,s. One pass is
2{,}500 steps of batch size 1 (about one epoch), AdamW with learning rate $10^{-4}$. After pass $n$ we compute the
validation loss at five fixed noise levels $\tau\in\{0.1,0.3,0.5,0.7,0.9\}$ over all validation windows,
\begin{equation}
  V_n = \frac{1}{5}\sum_{\tau} \frac{1}{|\mathcal{W}_{\mathrm{va}}|}\sum_{w\in\mathcal{W}_{\mathrm{va}}}
  \mathcal{L}_\tau(\theta_n; w),
\end{equation}
and stop at the first pass $n$ with $n - \arg\min_{k\le n} V_k \ge 3$, keeping
$\theta^\ast = \theta_{\arg\min_k V_k}$. Checkpoints every 500 steps make the run resumable.

\subsection{Inference and long playback}\label{sec:longplay}
Continuation keeps a 10\,s opening ($\mathbf{m}=1$ on its frames) and generates the next 25\,s; text-only generation
produces 35\,s from a description with $\mathbf{m}=0$. For long playback, segment $k$ is generated from the last
15\,s of the piece so far, $C_k$. Because generated clips end in a short silent tail, each segment is
over-generated by 3\,s and only the 30\,s that follow the context are kept; and because levels drift upwards from
segment to segment (Section~\ref{sec:chain}), the context is fed at a fixed level and the new part is matched to
the context:
\begin{equation}
  g_k = \frac{\rho}{\mathrm{RMS}(C_k)},\quad
  y_k = \frac{1}{g_k} G\!\left(g_k C_k\right)_{[15,\,45]\,\mathrm{s}},\quad
  \hat{y}_k = y_k \cdot \frac{\mathrm{RMS}(C_k)}{\mathrm{RMS}(y_k)},\quad \rho = 10^{-16/20},
\end{equation}
where $G$ denotes 48\,s generation conditioned on the given context; consecutive segments are joined with a 0.1\,s
crossfade.

\subsection{Baseline: a latent language model}
As an alternative to adapting a large generator, we trained a causal transformer from scratch that predicts the
next latent frame. With context noise augmentation \citep{pasini2024continuous},
$\tilde{\mathbf{z}}_{<i} = \mathbf{z}_{<i} + \sigma\boldsymbol{\eta}$ with $\sigma\sim\mathcal{U}(0,0.5)$ on half of
the samples, the transformer produces a state $\mathbf{h}_i = f_\phi(\tilde{\mathbf{z}}_{<i}, \sigma)$ and a
flow-matching head $u_\psi$ samples the frame \citep{li2024mar}:
\begin{equation}
  \mathcal{L}(\phi,\psi) = \mathbb{E}\left\| u_\psi\!\left((1-t)\boldsymbol{\epsilon} + t\,\mathbf{z}_i,\, t,\, \mathbf{h}_i\right) - (\mathbf{z}_i - \boldsymbol{\epsilon}) \right\|_2^2 .
\end{equation}
The model has 8 layers of width 512 and a 6-block AdaLN MLP head of width 1024 (72\,M parameters), and is trained
on 1.42\,M frames of the training split with sequences of 512 frames.

\subsection{Evaluation}\label{sec:eval}
All listening was done by a single professional listener with extensive experience of the guqin repertoire, on local web pages in
a fixed format: candidates for the same opening and seed are shuffled and unlabelled; each clip gets an overall
score from 1 to 5, optional issue tags (e.g.\ \textit{too many semitones}, \textit{sawing/metallic noise},
\textit{not guqin-like}, \textit{no melody}, \textit{abrupt seam}) and free-text comments; the mapping is revealed only
after rating. Openings come from compositions that the compared models did not train on, with different performers
and pieces across studies where possible. For paired score differences $d_1,\dots,d_n$ over openings we report the
exact two-sided sign-flip permutation test
\begin{equation}
  p = 2^{-n}\sum_{\mathbf{s}\in\{\pm1\}^n} \mathbf{1}\!\left[\left|\tfrac{1}{n}\textstyle\sum_i s_i d_i\right| \ge \left|\bar{d}\right|\right]
\end{equation}
and percentile bootstrap intervals of means; with eight or fewer pairs these tests have little power and we treat
them as descriptive.

Two automatic measures were tested against the ratings. The \emph{noisy-frame fraction} is the share of active
frames (within 45\,dB of the clip maximum) whose 2.5--10\,kHz band has spectral flatness above 0.25 and lies within
30\,dB of the 60--2500\,Hz band, aiming at the listener's ``sawing'' complaints. The \emph{pentatonic fit} $F$ of
Eq.~(\ref{eq:fit}) is computed on generated audio, and its onset-restricted variant counts the share of note onsets
whose attack pitch, estimated 30--120\,ms after a spectral-flux peak by harmonic summation
$S(n)=\sum_{q=1}^{5} 0.8^{\,q-1}\,|X(q f_n)|$ over semitones $n$, falls outside the opening's mode.

\section{Results}\label{sec:results}

\subsection{System performance}\label{sec:performance}
\begin{figure}[t]
  \centering
  \includegraphics[width=0.7\columnwidth]{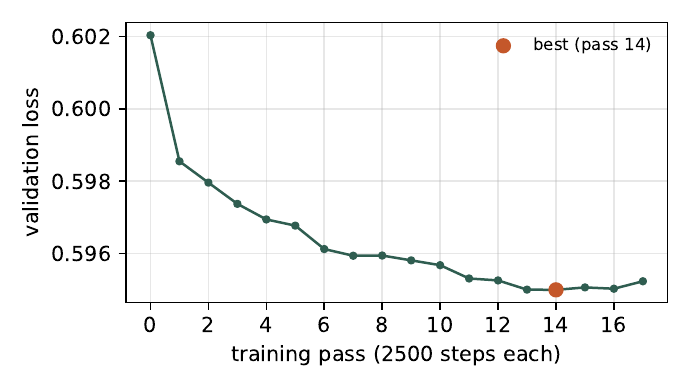}
  \caption{Validation loss per training pass of the final run (276 held-out windows, five fixed noise levels).}
  \label{fig:val}
\end{figure}

The final adapter is the pass-14 checkpoint of the 370-recording run (Figure~\ref{fig:val}): validation loss fell from
0.6020 for the unadapted model to 0.5950, and the three following passes did not improve on it. Training took
45.9\,min per pass (2{,}500 steps plus validation of 276 windows) on a single consumer GPU (NVIDIA RTX~5070~Ti, 16\,GB),
and 13\,h for the 17 passes. At inference the adapter generates 35\,s from text in 4.1\,s (8.6 times faster than real
time) and continues a 10\,s opening by 25\,s in 6.0\,s, with a peak of 6.5\,GB of GPU memory.

In the final listening study (R7 in Table~\ref{tab:studies}) the adapter scored a mean of \textbf{3.88} for
continuing eight unseen validation compositions, against 3.38 for the best previous adapter and 2.75 for an early
checkpoint of the same run, and \textbf{4.60} for five scene descriptions written after training, against 3.60 for
the early checkpoint. It received no ``sawing'' tags. One take (``sunset over a misty lake; an old fisherman rows
home, singing quietly to himself'') was singled out by the listener as exceptional and was extended to 3.5 minutes
by chained continuation (Section~\ref{sec:chain}).

\subsection{Listening studies}
Table~\ref{tab:studies} summarises the seven studies in the order they were run; each study answered a question
raised by the previous one.

\begin{table*}[t]
\centering
\small
\begin{tabular}{p{0.14\textwidth}p{0.47\textwidth}cp{0.23\textwidth}}
\hline
\colhead{Study} & \colhead{Condition} & \colhead{$n$} & \colhead{Mean score}\\
\hline
R1 masking & released default mask (10/80/10) / continuation mask (10/30/60) / 412 recordings, continuation mask & 8 & 2.88 / 2.62 / 3.38 \\
R2 inference & adapter strength 1.0 / 0.7 / adapter on early steps only & 9--10 & 3.56 / 3.60 / 3.67 \\
 & base model with 50 steps / earlier 222-clip adapter & 10 & 2.70 / 3.10 \\
R3 captions & continuation: no text in training / no prompt / matching / opposite tags & 9 & 3.89 / 3.56 / 4.00 / 4.00 \\
 & text only: adapter trained without / with captions & 5 & 2.00 / 3.80 \\
R4 latent LM & from-scratch latent LM / SA3 adapter & 11 / 6 & 1.00 / 2.00 \\
R5 length & 1 pass / 3 passes + tags / 3 passes + tags + mode & 8 & 3.75 / 3.75 / 4.12 \\
R6 narratives & pass 2 / adapter C from R5 & 7 / 8 & 3.00 / 3.50 \\
R7 final & continuation: pass 14 / pass 2 / adapter C & 8 & \textbf{3.88} / 2.75 / 3.38 \\
 & scene descriptions: pass 14 / pass 2 & 5 & \textbf{4.60} / 3.60 \\
\hline
\end{tabular}
\caption{Overview of the listening studies (scores 1--5, single expert listener). In R1 the 412-recording adapter
had seen the validation compositions; all other comparisons use openings unseen by every compared model, except R2
(see the limitations in Section~\ref{sec:discussion}).}
\label{tab:studies}
\end{table*}

\paragraph{Masking and data (R1).} With 329 training recordings, the continuation-weighted mask did not beat the
released default mixture (2.62 against 2.88), and the two seeds differed by 1.6 points on average (2.17 against 3.75), more than any
model difference.

\paragraph{Inference settings (R2).} Lowering adapter strength or restricting the adapter to the early, high-noise
steps changed the mean by at most 0.1. Running the base model for 50 steps, the configuration that matches training,
was clearly worse (2.70, five scores of 1, comments about extra instruments). The mean per opening-and-seed group
ranged from 1.0 to 4.75.

\paragraph{Text conditioning (R3).} Training with mood tags left continuation unchanged (matching and deliberately
opposite tags both scored 4.0) but improved generation from text alone from 2.0 to 3.8 ($p=0.06$, $n=5$).

\begin{figure}[t]
  \centering
  \includegraphics[width=0.7\columnwidth]{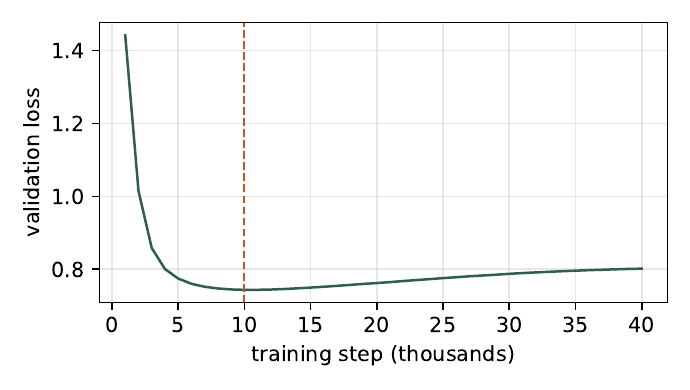}
  \caption{Validation loss of the from-scratch latent language model; it rises after 10k steps (dashed line).}
  \label{fig:lm}
\end{figure}

\paragraph{From-scratch latent language model (R4).} Validation loss of the baseline reached its minimum at 10k steps
and rose by 8\% by 40k steps (Figure~\ref{fig:lm}). All 11 of its clips, including continuations, 5-minute endless
generation and unconditioned generation, received the lowest score, with comments such as ``no timbre at all''; the
SA3 adapter scored 2.0 on the same openings.

\paragraph{Training length and mode (R5, R6, R7).} On the 329-recording split, held-out loss decreased monotonically
over three passes (0.5387 for the base model, then 0.5351, 0.5343 and 0.5337), and neither tags nor mode text changed
it. Listening favoured three passes with tags and mode (4.12 against 3.75 for one pass, $p=0.50$). The story-caption
run was rated below that adapter after two passes (3.00 against 3.50) but above it after fourteen (3.88 against 3.38,
$p=0.63$; against pass~2, $p=0.16$). Bootstrap 95\% intervals of these means span about two points (pass~14:
[2.75, 4.75]).

\paragraph{Automatic measures.} The noisy-frame fraction correlated with the score in the wrong direction (Spearman
$\rho=+0.39$ over 48 clips), and the nine clips tagged as sawing had \emph{lower} values than the rest (0.0006 against
0.0017). The pentatonic fit tracked scores over 124 rated clips ($\rho=0.52$, positive in each study) but only weakly
among candidates for the same opening ($\rho=0.31$ after removing group means); choosing the best-fitting candidate
raised the mean from 3.41 to 3.52, discarding the worst raised it to 3.58. The onset-restricted variant was weaker
still ($\rho=0.19$ within groups), although generated clips did place more onsets outside the mode than real
recordings (33\% against a median of 22\%; 39\% in clips flagged for semitones).

\paragraph{Rater consistency.} Fifteen clips from R6 were rated again, unknowingly, in R7: six scores were identical,
twelve within one point, and the mean absolute difference was 0.8.

\subsection{Long playback by chained continuation}\label{sec:chain}
Extending the favourite take exposed two problems. Every generated clip ends with about 0.5\,s of digital silence
($-87$\,dBFS) as the requested duration runs out; fed back as context, the tail made the model insert a further two
seconds of silence at every seam. And each segment came out slightly louder than its context, so that over six
rounds the level rose from $-10$ to $-5.5$\,dBFS with up to 10\% of samples clipped. Tail trimming and level locking
(Section~\ref{sec:longplay}) removed the gaps and kept a 3.5-minute piece between $-13.8$ and $-16.7$\,dBFS per 30\,s
with no clipping. Of two locked chains the listener accepted one and rejected the other for a sudden breakdown after
82\,s.

\section{Discussion}\label{sec:discussion}

\paragraph{Randomness dominates; selection is part of the system.} In every study the opening and the random seed
moved scores more than any setting we varied. The same semitone often appeared at the same second in all candidates
sharing an opening and seed, which locates the error in the sampled trajectory rather than in a particular adapter.
For a radio this has a practical consequence: generating several takes and choosing among them is not an
afterthought but a component of the system, and the quality of that choice is bounded by the quality of the
selection criterion, which we did not yet find automatically.

\paragraph{The pretrained prior carries the timbre.} The from-scratch latent language model produced no recognisable
guqin at all, even though the same decoder renders guqin well from encoded recordings. Sampling 256-dimensional frames
one at a time from 1.4\,M training frames left the trajectory off the region of latent space that the decoder renders
as guqin; noise augmentation of the context did not prevent it, and its validation loss began to rise after 10k of
40k training steps. Adapting a generator pretrained on far more audio, by contrast, kept timbre largely intact from
the first adapter on: the listener's complaints were about melody and noise, not about the instrument. In a domain of
tens of hours, the question is not which architecture to train but which prior to adapt. A related lesson concerns the
two Stable Audio~3 models: the base model, which matches the training configuration of the adapter, rendered worse
than the distilled model, which the adapter was never trained on.

\paragraph{What text does and does not control.} Captions had no measurable effect on held-out loss and none on
continuation, where the ten-second opening determines key, register and texture. They mattered only when there was no
opening: text-only generation improved from 2.0 to 3.8 after training with mood tags, and the narrative captions
produced the highest-rated pieces of the project from scene descriptions that appear nowhere in training. Narratives
attached to one composition each risk being learned as composition identifiers, whereas tags shared across
compositions force the model to associate words with sound; mixing both, as we did, keeps the shared vocabulary
while letting free text reach the encoder.

\paragraph{Why the automatic measures failed.} The friction noise of slides is part of the guqin's sound, which
\cite{penttinen2006guqin} model as a separate excitation component; what made the listener hear ``sawing'' was not its
amount but where it occurred and how it was articulated. A frame-level noise statistic cannot see that, and it
correlated with the score in the wrong direction because noisier clips were often livelier ones. The pentatonic fit
captured a real property, and generated clips did place more attacks outside the mode than real recordings, but the
listener's own description, that idiomatic semitones occur inside slides rather than at the pluck, calls for
technique-aware analysis of the kind developed for playing-technique detection
\citep{huang2020guqin,li2022guzheng}. The measure also disagreed with the listener about the effect of longer
training, which is the strongest warning against using it to make decisions. We therefore treat both measures as
negative results and use the listener's ratings as the only criterion.

\paragraph{Training longer helped, but loss is a weak signal.} Held-out loss decreased by only 1.2\% from the unadapted
model to the best pass, and the differences between late passes were of the order of $10^{-5}$, yet the pass-14
adapter was the best-rated model of the project. Most of the diffusion loss is irreducible noise, so a small and
monotone decrease can coincide with a large perceptual change. Loss is nevertheless the right instrument for one
question, when to stop, because it is measured on held-out compositions; it would have been tempting, and wrong, to
stop at the first noisy uptick (passes 8 and 15) instead of waiting for three passes without a new best.

\paragraph{Long playback is an engineering problem first.} The two failures of chained continuation, silent tails and
loudness feedback, are properties of how a fixed-length generator ends and of how its output level relates to its
context, not of musical modelling, and both disappeared with a few lines of signal processing. The remaining failure,
a sudden breakdown after 82\,s in one of two chains, is musical and is the real obstacle to hours of playback; it is
the natural target for selection between candidate continuations and for preference-based fine-tuning on the
collected ratings \citep{wallace2023diffusiondpo}.

\paragraph{Evaluation with one expert.} A single guqin-literate listener made fast iteration possible and produced
detailed, actionable comments (``the hook repeats the original'', ``semitones like a beginner practising'',
``sounds like a cowboy guitar''), several of which turned directly into method changes. The cost is reliability:
repeated ratings of the same clips differed by 0.8 points on average and a showcase take was rated 1 in a later
session, effects as large as the differences between our models. None of our comparisons is significant, and the
studies were sequential, so later choices depended on earlier ratings. Multi-listener evaluation is the first
requirement for stronger claims.

\paragraph{Data curation was the decisive step.} Automatic screening with audio tagging and audio-text models served
only as a first pass; the final decision on every recording was made by whole-track listening. Composition-level
splits, a check of a 10-hour DVD collection against the CD library (which turned out to be
mostly new but full of applause and ensemble playing), and the discovery that openly licensed guqin audio amounts to
under an hour all shaped what could be trained. A planned held-out performer was, by mistake, included in the
412-recording training set, which compromised one study; keeping held-out material out of every automated corpus
build is worth an explicit check.

\paragraph{Limitations.}\label{sec:limits} All ratings come from one listener; the studies are small and were
run sequentially as development proceeded. In R2 the openings came from a performer who, contrary to the original
plan, had been included in the 412-recording training set. The narratives and tags are condensed readings of
traditional notes on each piece and simplify a rich interpretive tradition; 22 of the 108 titles are inferred from
the title alone. We did not test for replication of training material \citep{batlleroca2024replication}, and the
pitch and noise measures were validated only against the same listener.

\section{Conclusion}\label{sec:conclusion}

A pretrained latent diffusion model, 42.5\,h of carefully screened historical recordings, a 21.6\,M-parameter adapter
and a single attentive listener were enough to produce guqin continuations and scene-driven pieces that the listener
rated highly, at more than eight times real time on a consumer GPU, while training a generator from scratch on the
same data failed completely. The strongest effects came not from architecture but from data curation, the choice of
prior, held-out early stopping and simple engineering of the generation loop. The next steps are multi-listener
evaluation, technique-aware selection between candidate takes, preference-based fine-tuning on the collected ratings,
and stable long-form generation for continuous playback.

\section{Reproducibility}

Code for corpus preparation, training, generation, listening pages and analysis, together with every listening
rating and validation log and the scripts that produce all numbers and figures in this article, is available
\ifdefined\anonymous in an anonymised repository (link withheld for review)\else at
\url{https://github.com/HuanchenCai/guqin-generation} (tag \texttt{arxiv-v1}; results were frozen at tag
\texttt{freeze-2026-10-05}, whose \texttt{FREEZE.json} records the SHA-256 hashes of all checkpoints used)\fi. The training recordings come from a personal collection of
historical performances that remain under copyright and cannot be redistributed, and for the same reason we do not release audio
generated from them or the adapter weights. The base model is used under Stability AI's community license.

\section{Ethics and Consent}

The listening studies involved a single expert listener. No personal data were collected.

\section*{Acknowledgements}
\ackText

\section*{Competing Interests}
The author has no competing interests to declare.

\section*{Author Contributions}
\contribText

\section*{Use of AI Tools}
\aiText

\bibliographystyle{apaTISMIR}
\bibliography{references}

\end{document}